\documentstyle[12pt,aps]{revtex}
\begin{document}
\title{
Chiral symmetry and properties of hadron correlators in matter\bigskip}
\author{ Boris Krippa \thanks{on leave from the 
Institute for Nuclear Research of
 the Russian Academy of Sciences, Moscow Region 117312,
Russia.}} 
\address{Department of Physics, Free University of Amsterdam,
De Boelelaan 1081, 1081 HV Amsterdam, the Netherlands. \\}
\maketitle
\bigskip
\vspace{-5.5cm}
\vspace{5.5cm}
\begin{abstract}
The constraints imposed by chiral symmetry on hadron correlation
functions in  nuclear medium are discussed.
It is shown that these constraints
imply a certain structure for the in-medium hadron correlators
and lead to the cancelation of the order $\rho m_\pi$  term
in the in-medium nucleon correlator. We also consider the effect
of mixing of the chiral partners correlation functions  
arising from the interaction
of nuclear pions with corresponding interpolating currents.
It reflects the phenomena of partial restoration of chiral symmetry.
The different scenarios of such restoration are briefly discussed.
\end{abstract}
\vskip0.5cm
Keywords: chiral symmetry, mixing effect, hadron correlators, pions
\vskip0.2cm
PACS: 11.30.Rd, 11.55.Hx, 21.65.+f
\newpage
Chiral symmetry (CS) has become  one of the most important principles
of  modern nuclear physics \cite {Ad}. In the limit
 of massless quarks the QCD 
Lagrangian is symmetric with respect to the transformations belonging
to the $SU(N) \times SU(N)$ chiral group. It is generally believed
that this symmetry is spontaneously broken so that the ground state
of the theory does not enjoy the symmetry of the underlying Lagrangian.
In the case of the $SU(2) \times SU(2)$ group, which is of  relevance 
for nuclear physics, the Goldstone theorem implies the existence
of 3 massless bosons which are responsible for the 
$NN$ interactions. CS breaking manifests itself in the absence
of chiral multiplets of the particles with the same masses but
different parities, for example, $\rho-a_1$ or $\sigma-\pi$
mesons. In the language of  correlation functions the broken chiral 
symmetry means that the lowest pole positions of vector and axial 
vector correlators, describing the $\rho$ and $a_1$ mesons
, are different. Thus,  restoration of the symmetry in vacuum would
 result in the identity of the corresponding
correlators. It  in turn leads to the same masses of the chiral partners.
One notes, however that restoration of CS can happen in the academic limit
 when the number of the light flavors
is more than 5 \cite{Sm}. 
In matter the pattern of complete or partial restoration of CS
is more complicated, as discussed below. The other manifestation of the
 spontaneously broken CS is the occurrence of  nonzero order parameters.
One of the possible order parameters is the expectation value of the two
quark condensate ${\overline qq}$. It is believed in the standard scenario
of hadron interactions in vacuum that $q{\overline {q}}$=0 would imply the
restoration of CS and therefore the masses of hadrons, the larger
part of which is due to 
CS breaking, should become much smaller compared to their observed values
(we put aside more exotic possibilities of having
spontaneously broken CS with significantly reduced or even vanishing
two quark condensate \cite{St}).
 One notes that the relationships
between condensates, hadron masses and corresponding correlators  
are significantly more complicated in the presence of medium.
For example, in the lukewarm pion gas the temperature dependence
of the two quark condensate ${\overline {q}q}(T)$ and the in-medium nucleon
mass is different \cite{El} so that the change of the nucleon mass in
the medium is not completely determined by the corresponding change of 
the quark condensate. An analogous conjecture is valid in nuclear matter
\cite {Birse}. We will discuss this latter case in more detail below.
The other example is the issue of
chiral symmetry restoration for the correlators of the chiral partners. 
In this case the restoration of the symmetry which means 
 the identity of these correlators, in vacuum would 
 imply the equality of the masses of chiral partners. The
 presence of matter makes things more complicated and the identity of
the corresponding in-medium correlators does not necessarily mean the
degeneracy of the effective masses of chiral partners. The phenomena
of axial-vector mixing established both at finite temperature \cite{De}
and  density \cite{Kr} provides such an example. This mixing naturally
leads to  several possible scenarios of CS restoration in matter.
The identity of the masses of chiral partners at the point of restoration
is only one of them. In this paper we address the issue of constraints
which chiral symmetry imposes on the hadron correlators in nuclear medium.
The difference between the cases of finite density and temperature
is that in nuclear medium pions may induce the formation of the particle-hole
excitations in contract to the thermal case where creation or absorption
of pion does not change the characteristics of a heat bath.
  
Some of the results presented in this paper were earlier discussed in
short letters \cite {Kr,Bk}. Here we present the formal aspects of
calculations in greater details,
 elaborate some other properties of the hadron correlators following
from the the CS constraints  and give an extended discussion of the
 problems involved. We derive a new scalar-isoscalar type of chiral mixing 
and discuss its physical consequences.  The chiral 
corrections to the lowest order in-medium scalar two-quark condensate
are also calculated. Finally we extend the formalism to the case of 
3-point functions.
The paper is organized as follows.
In the next section we derive  the leading chiral 
corrections for the in-medium hadron correlators. In the section 3
we consider the case of the nucleon correlators and discuss the relationships
between density dependence of the two-quark condensate and nucleon mass.
The next section is devoted to the phenomena of the mixing between the 
vector and axial-vector correlators
 and possible scenarios of CS  restoration.
In the section 5 we  discuss the leading chiral
corrections for the two-quark
condensate and in the next two sections we consider the effect of chiral mixing
in the scalar-pseudoscalar channel and discuss the general case of 
3-point functions.
 We summarize and conclude in the last section.

{\bf 2.} The dynamics of hadrons in nuclear medium is
 described by the corresponding
correlators. Let's consider the case of the two-point correlators. They can 
be written in the form    
\begin{equation}
\Pi(p)=
i\!\int d^4x e^{ip\cdot x}\langle \Psi|T \{J(x)J(0)\}
|\Psi\rangle.
\end{equation}
Where $J$ is the interpolating current in the corresponding hadron channel
and the matrix element is taken over the ground states of the system with
finite density $\rho$. In each particular case the interpolating current
should be supplied with the corresponding Lorenz and isospin indices to 
specify the hadron channel considered. The position of the lowest pole and
its behavior as the function of density determines the in-medium mass 
of hadron. The medium effects are included in the nuclear wave function
$\Psi$. If this wave function describes an infinite system of non 
interacting nucleons than the above correlator reflects the dynamics
of the hadronic probe  moving in some mean field formed by nuclear matter.
To calculate the corrections to this rather crude picture one needs to 
take into account the nuclear pions which can be absorbed or emitted by
nucleons and provide the interactions among them. It is convenient to use
the Fock representation of the nuclear wave function.
\begin{equation}
\Psi = \Psi^A|A\rangle + \sum_{a}\Psi^A_a |A\pi_a\rangle +
  \sum_{a,b}\Psi^A_{a,b} |A\pi_{a}\pi_{b}\rangle
+ .......
\end{equation}
Here $A$ represents the state vector in the Fock space describing the nuclear
system of $A$  nucleons without pions,  $|A\pi_{a}\rangle$
is the Fock state of the nuclear system with $A$ nucleons plus one pion
and  $|A\pi_{a}\pi_{b}\rangle$ denotes 
the Fock state with $A$ nucleons
and two pions. The summation goes over the corresponding pion quantum numbers.
We truncated the series taking into account the states with zero, one and
two pions since the main contribution and leading chiral corrections are given
by the matrix elements taken with these states. 
The states with larger number of pions would give  corrections of 
higher chiral dimension. Since the nuclear pions are virtual, all 
possible time orderings should be taken into account. Putting the above
expression for the nuclear wave function into Eq.(1) and making use of the
identities 
\begin{equation}
\langle\Psi^A|\Psi^{A}_{a,b}\rangle = 
\langle\Psi^A|a^{+}_{a}a^{+}_{b}|\Psi^A\rangle , 
\end{equation}
\begin{equation}
\langle\Psi^{A}_{a,b}|\Psi^A\rangle = 
\langle\Psi^A|a_{a}a_{b}|\Psi^A\rangle , 
\end{equation}
and
\begin{equation}
\langle\Psi^{A}_{a}|\Psi^{A}_{b} \rangle = 
\langle\Psi^{A}|a_{a}a_{b}^{+}|\Psi^A\rangle  
\end{equation}
one can represent the in-medium hadron correlator by the sum of two pieces
\begin{equation}
\Pi(q) = \overcirc\Pi(q) + \Pi^{\pi}(q)
\end{equation}
Here the first term corresponds to the contributions from the system of
noninteracting nucleons and can be written as follows
\begin{equation}
\overcirc {\Pi}(q)=i\!\int d^4x e^{ip\cdot x}\langle A|T \{J(x)J(0)\}
|A\rangle, 
\end{equation}
whereas the second one describes the pionic corrections when the
interpolating current interacts directly with the nuclear pions.
Let's consider the specific part of these corrections  where the pion is first
emitted and then absorbed by nuclear matter. The corresponding piece of
the correlators can be represented in the form
\begin{equation}
\sum_{a,b}\int {d^3\hbox{\bf k}\over 2\omega_k}
{d^3\hbox{\bf k}'\over 2\omega_k'}\langle\Psi^A |a^{a\dagger}(\hbox{\bf k})
a^b(\hbox{\bf k}')|\Psi^A\rangle
\,i\!\int d^4x e^{ip\cdot x}\langle A\,\pi^a(\hbox{\bf k})|T\{J(x)
J(0)\}|A\,\pi^b(\hbox{\bf k}')\rangle,
\label{piinout}
\end{equation}
As we mentioned earlier, due to virtual character of the nuclear pions
all time orderings allowing pions to go backward or forward in time should
be taken into account. These pieces of the correlator
$\Pi^{\pi}(q)$ can be written as follows
\begin{equation}
{1\over 2}\sum_{a,b}\int {d^3\hbox{\bf k}\over 2\omega_k}
{d^3\hbox{\bf k}'\over 2\omega_k'}\langle\Psi^A|a^a(\hbox{\bf k})
a^b(\hbox{\bf k}')|\Psi^A\rangle
\,i\!\int d^4x e^{ip\cdot x}\langle A|T\{J(x)J(0)\}
|A\,\pi^a(\hbox{\bf k})\pi^b(\hbox{\bf k}')\rangle,
\label{twopiin}
\end{equation}
and
\begin{equation}
{1\over 2}\sum_{a,b}\int {d^3\hbox{\bf k}\over 2\omega_k}
{d^3\hbox{\bf k}'\over 2\omega_{k'}}\langle\Psi^A|a^{a\dagger}(\hbox{\bf k})
a^{b\dagger}(\hbox{\bf k}')
|\Psi^A\rangle
\,i\!\int d^4x e^{ip\cdot x}\langle A\,\pi^a(\hbox{\bf k})\pi^b(\hbox{\bf 
k}')|
T\{J(x)J(0)\}|A\rangle
\end{equation}
where $\omega_k=\sqrt{{k}^2+m_\pi^2}$.
One notes that the correlator $\overcirc\Pi(p)$ with noninteracting
 nucleons
was used for calculating of the in-medium properties of nucleon 
\cite{Dr,Co} and $\rho$-meson \cite{Ha} in the framework of QCD sum rules.
We consider the nuclear pions in the chiral limit. Since we are interested
in the properties of hadron correlators which are related to chiral symmetry 
treating nuclear pions in the chiral limit seems to be a reasonable 
assumption in our case.  Moreover, if one believes that the hadron mass
generation mechanism is due to spontaneous breaking of CS
then a finite pion mass would act as a small perturbation. This conjecture is 
partly supported by the results of Ref. \cite{Bu} where it was shown that
in the chiral limit nuclear properties  (at least the static ones)  
experience little change compared to the real case. 
One notes that the validity of using chiral limit for nuclear pions strongly
depend on the concrete problem under consideration. If one is interested in 
the properties of correlators which are related to the phenomena of 
spontaneously broken chiral symmetry such as chiral mixing or behavior of the 
two-quark condensate as a function of density  than treating nuclear pions
 in chiral limit is a reasonable approximation since in this case the physical 
pion mass should act as a small perturbation. If the problem under study is, 
for example, to compute the widths and energy levels of the pionic mesoatom or
calculate pion-nucleus reaction amplitude than the physical pion mass should be
taken into account.
 
The use of the soft-pion
theorem gives 
\begin{equation}
i\langle A\,\pi^a(\hbox{\bf k})|T\{J(x),
J(0)\}|A\,\pi^b(\hbox{\bf k}')\rangle
\simeq {-i\over f_\pi^2}\langle A|\left[Q_5^a,
\left[Q_5^b, T\{J(x),J(0)\}\right]\right]|A\rangle,
\label{soft}
\end{equation}
The other matrix elements can also be reduced to similar expressions
in the same manner. Now the part of the correlator describing
the pionic corrections can be written
as follows
\begin{eqnarray}
\Pi^{\pi}={-i\over 2f_\pi^2}\sum_{a,b}\int {d^3\hbox{\bf k}\over 2\omega_k}
{d^3\hbox{\bf k}'\over 2\omega_{k'}}\langle\Psi^A|((a^{+}_a(\hbox{\bf k})
a^{+}_b(\hbox{\bf k}')+a_a(\hbox{\bf k})
a_b(\hbox{\bf k}')+\nonumber\\
2a^{+}_a(\hbox{\bf k})
a_b(\hbox{\bf k}'))|\Psi^A\rangle
\,i\!\int d^4x e^{ip\cdot x}
\langle A|[Q^{a}_{5},[Q^{b
}_{5},T\{J(x),J(0)\}]]|A\rangle,
\end{eqnarray}
The first integral in this expression  is proportional to the result of 
expansion of the matrix element
$\langle \Psi^A|{1\over 2}\pi^2(0)|\Psi^A\rangle$ in terms of creation
and annihilation operators. Here $\pi(0)$ is the pion field operator and
 normal ordering is assumed. To the first order in density this matrix
element is 
\begin{equation}
(2\pi)^3\langle \Psi^A|{1\over 2}\pi^2(0)|\Psi^A\rangle m_\pi^2
\simeq (2\pi)^{3} m_\pi^2 \langle N|\pi^{2}(0)|N\rangle\rho \simeq 
\rho\overline{\sigma}_{\pi{\scriptscriptstyle N}}
\end{equation}
In this expression the matrix element is taken over the nucleon states
and $\overline{\sigma}_{\pi{\scriptscriptstyle N}}$
is the leading nonanalytic  part of the pion-nucleon sigma term
$\sigma_{\pi{\scriptscriptstyle N}}$. The appearance of this nonanalytic
 term 
 can qualitatively be understood as follows.
The chiral expansion of $\sigma_{\pi{\scriptscriptstyle N}}$ reads as
\begin{equation}
{\sigma}_{\pi{\scriptscriptstyle N}}=A m_\pi^2 -
 {9\over 16}({g_{\pi N}\over 2M_N})^2m_\pi^3 + ....... 
\end{equation}
The first, analytic term is proportional to some constant which contains a
counterterm needed to regularize the short range contributions.
In contrast, the nonanalytic term is due to the long distance contribution
of the pion cloud. Being long-ranged this term is governed by CS
 and thus should enter the expression describing the leading
chiral corrections, where the main role is played by the nuclear pion
cloud. The actual type of the chiral expansion of the in-medium
hadron correlator depends on the commutation relation between the
chiral generator $Q^a_5$ and the  hadron interpolating current chosen.
One notes that the factorization, considered above is valid
in the soft pion limit. The entire nuclear pion cloud is peaked
around 400MeV/c. However, since we deal with effects due to chiral symmetry
we are interested in the soft part of this cloud so the chiral 
 limit is relevant. To include the corrections to 
this limit  one
needs to consider the effects which may not be governed by chiral symmetry.
This is beyond the scope of this paper. It is important to note that chiral
corrections related to the nuclear pions may, in principle, be accompanied
by the high lying excitations of the $2p-2h$ states, to ensure isospin
conservation. Of course, the wave functions
   of the exited 2p-2h and ground states are very different but
the difference between the corresponding matrix elements is presumably of higher chiral order.
One notes that the excitation of the   2p-2h  states is in some sense a pure ``nuclear'' effect. 
From the chiral  symmetry point of view the nuclear states without pions and with one pion are related
by chiral rotations so the matrix elements like $<2p-2h|O|2p-2h>$ and $<ground|O|ground>$ should be the
 same in leading chiral order. It can qualitatively be understood as follows.
 The chiral scale is higher than the nuclear one so the matrix elements taken between the 
states, which can be chirally rotated to each other but correspond to the different eigenvalues of
 the nuclear Hamiltonian should be the same in leading order at the chiral scale and can differ
only by a factor of order the typical nuclear scale. Since nuclear scale is significantly lower
than the chiral one the difference between matrix elements should show up at higher order of chiral
 expansion, but not at the leading one.

In the next section we consider the case where the interpolating current
relates vacuum and the states with  nucleon quantum numbers.

{\bf 3.} The renormalization of nucleon mass in dense matter has been one of the
main issues in nuclear physics for a long time. Many recent
studies of the in-medium nucleon dynamics have to large extent been focused
on the relations between the phenomena of partial restoration of CS
and changes of nucleon properties in matter. As we already mentioned,  
this hidden symmetry is characterized by some order parameter which in 
our case is provided by the nonzero vacuum expectation value of the two 
quark condensate $\overline{q}q$. Matter can influence the properties of 
the QCD vacuum and thus change the two quark condensate. It is usually
believed that, in general, the reduction of  $<\overline{q}q>$ is related
in some way to the effect of partial restoration of chiral symmetry. 
To the first order in the density the evolution of the two-quark condensate
is given by \cite{Dr,Co}
\begin{equation}
\langle\Psi|\overline {q}q|\Psi\rangle=\langle 0|\overline {q}q|0\rangle
(1-{\sigma_{\pi{\scriptscriptstyle N}}\over f_\pi^2 m_\pi^2}\rho),
\label{ratio}
\end{equation}
where $|\Psi\rangle$ denotes our nuclear matter ground state.
Much attention has recently been paid to the issue of 
the higher density corrections to this
result \cite{We,Ko}. At present there is  qualitative agreement that the 
contributions of the higher order terms are moderate  up to
the normal nuclear matter density. On the contrary, there have been much less
discussions on how the changes of the two quark condensate affect the 
in-medium particle properties and whether any change of  
$\overline qq$ is in  one-to-one correspondence with the  
CS restoration phenomena. This problem was first discussed
in ref.\cite{Birse}. The reasoning was the following. One considers the 
scalar self-energy of a nucleon in nuclear matter. Let this
 self-energy depend in 
some way on the quark condensate. Then one should be a piece in the
 the self-energy proportional to the second term in the
 above written expression for the in-medium quark condensate. At moderate
densities the nucleon self-energy can be represented by the product of
the scalar part of the $NN$
amplitude and nuclear density $\rho$ so there
will be a piece of the  $NN$ amplitude which is proportional to
 $\sigma_{\pi\scriptscriptstyle N}\over m_\pi^2$. Making use of the chiral
 expansion of the pion-nucleon sigma term one can easily see the the $NN$
amplitude contains a term of order $ m_\pi$  not allowed by chiral
counting rules \cite{Wein} according to which there is no term of this
order in the chiral expansion of the $NN$ amplitude. Thus one can conclude
that there is no direct correspondence between changes of the quark condensate
in nuclear matter and the in-medium nucleon properties. One notes that
this conclusion is in complete accord with the results obtained by Cohen et al.
\cite {Lee} where it was pointed out that, in order to be 
consistent with chiral symmetry, the expression for the 
nucleon mass in vacuum should contain no  terms of order
$m_q$ times by the  logarithm of the quark mass
 whereas these terms are present in the quark condensate.
It was demonstrated in \cite{Lee} how, using  QCD sum rules, one can
remove the unwanted pieces. Technically such terms in the Operator Product
Expansion (OPE) can be canceled by the corresponding piece of the 
phenomenological side of QCD sum rules provided a careful treatment of
the continuum, including the low-momentum $\pi N$ states, is made. As we show 
below, similar cancelation takes place in nuclear matter. It is again
convenient to use the QCD sum rule approach to demonstrate 
how this  cancelation appears in the case of nuclear matter. In QCD
sum rules one relates the characteristics of  QCD vacuum and the 
phenomenological in-medium nucleon spectral density. The OPE consists of
certain set of local operators responsible mainly for short range effects.
The effects of the low-momentum pions are long-range and should, therefore,
stem from the phenomenological
 representation of the
 in-medium nucleon correlator. The correlator of two nucleon interpolating
currents can be written as follows     
\begin{equation}
\Pi(p)=i\!\int d^4x e^{ip\cdot x}\langle \Psi|T\{\eta(x)\,\overline{\eta}(0)\}
|\Psi\rangle.
\end{equation}
Assuming the Ioffe choice \cite{Iof} of the nucleon
interpolating current, making use
 of the transformation property of this current 
\begin{equation}
\left[Q_5^a, \eta(x)\right]=-\gamma_5{\tau^a\over 2}\eta(x).
\label{fchtr}
\end{equation}
one can get the following chiral expansion of the in-medium nucleon 
correlator
\begin{equation}
\Pi(p)\simeq\overcirc\Pi(p)-{\xi\over 2}\Bigl(\overcirc\Pi(p)+\gamma_5 
\overcirc\Pi(p)\gamma_5\Bigr),
\label{correxp}
\end{equation}
where we defined 
\begin{equation}
 \xi={\rho\overline{\sigma}_{\pi{\scriptscriptstyle N}}\over f_\pi^2 m_\pi^2}
\end{equation}
and $\overcirc\Pi(p)$ is the nucleon correlator in the chiral limit.
It is useful to to decompose the correlator into three terms with different
Dirac structures \cite{Co}
\begin{equation}
\Pi(p)=\Pi^{(s)}(p)+\Pi^{(p)}(p)p\llap/+\Pi^{(u)}(p)u\llap/,
\end{equation}
where $u^\mu$ is a unit four-vector  defining the rest-frame of nuclear
system.
One can see that only the piece $\Pi^{(s)}(p)$ gets affected by the chiral
corrections of order $\rho m_\pi$. Given the fact that the OPE for  $\Pi^{(s)}(p)$
involves terms transforming like two-quark condensate $\overline {q}q$
whereas $\Pi^{u(p)}(p)$ contains chirally invariant matrix elements such as
$G_{\mu\nu}G^{\mu\nu}$ or $q^{+}q$ this result looks quite natural.
Splitting, as  is usually done in the QCD sum rule method, 
the phenomenological expression of the nucleon correlator into pole and
 continuum parts one can obtain
\begin{equation}
\Pi(p)\simeq\Pi_{pole}(p)-{\xi\over 2}\gamma_5\Pi_{pole}(p)\gamma_5
+\left(1-{\xi\over 2}\right)\overcirc\Pi_{cont}(p)
-{\xi\over 2}\gamma_5 \overcirc\Pi_{cont}(p)\gamma_5,
\label{corrphen}
\end{equation}
Where we denoted $\Pi_{pole}(p)\simeq (1-{\xi\over 2})\overcirc\Pi_{pole}(p)$.
The explicit expression of the pole term has the following form
\cite{Co}
\begin{equation}
\Pi_{pole}(p)=-\lambda^{*2}{p\llap/+M^*+V\gamma_0\over 2E(\hbox{\bf p})
[p^0-E(\hbox{\bf p})]},
\label{pole}
\end{equation}
Here
$M^*$ is the in-medium nucleon mass including the scalar part of the self
energy and  $\lambda^*$ is the coupling of the
nucleon interpolating current
with the corresponding lowest state.
According to the standard ideology of QCD sum rules the OPE and
 phenomenological sides should be matched with some weighting functions.
Then one writes the three independent sum rules, one for each Dirac structure
\begin{eqnarray}
-\left(1-{\xi\over2}\right)\lambda^{*2}M^*\int d^4p\,{w(p)\over 
2E(\hbox{\bf p})[p^0-E(\hbox{\bf p})]}&\simeq&(1-\xi)\int d^4p\, w(p)\left[
\overcirc\Pi^{(s)}_{OPE}(p)-\overcirc\Pi^{(s)}_{cont}(p)\right],\nonumber\\
-\left(1+{\xi\over2}\right)\lambda^{*2}\int d^4p\,{w(p)\over 2E(\hbox{\bf p})
[p^0-E(\hbox{\bf p})]}&\simeq&\int d^4p\, w(p)\left[\overcirc\Pi^{(p)}_{OPE}(p)
-\overcirc\Pi^{(p)}_{cont}(p)\right],\label{sumrules}\\
-\left(1+{\xi\over2}\right)\lambda^{*2}V\int d^4p\,{w(p)\over 2E(\hbox{\bf p})
[p^0-E(\hbox{\bf p})]}&\simeq&\int d^4p\, w(p)\left[\overcirc\Pi^{(u)}_{OPE}(p)
-\overcirc\Pi^{(u)}_{cont}(p)\right].
\end{eqnarray}
Taking the ratio of these sum rules one can get the needed cancelation
in the effective mass and vector self energy and bring the in-medium
nucleon QCD sum rules into agreement with the chiral symmetry constraints. 
The numerical value of $\overline{\sigma}_{\pi{\scriptscriptstyle N}}$ is
 about -20MeV to be compared with the numerical 
value of the entire pion-nucleon sigma-term 
$\sigma_{\pi{\scriptscriptstyle N}}$ = 45MeV, so that the numerical
effect is quite significant. One notes that from the chiral expansion
of the nucleon correlator one can easily see that the factorization 
approximation, sometimes used 
for the four quark condensate, cannot be valid in matter since in 
does not satisfy the chiral symmetry requirements. This condensate enters
the OPE of $\Pi^{p(u)}(p)$ which, being chiral invariant, is not affected by
chiral corrections. When  factorization is used the chirally noninvariant
product of two two-quark condensates appears in the OPE of  $\Pi^{p(u)}(p)$
bringing in the terms of order $\rho m_\pi$ which are inconsistent with chiral
symmetry.\\
{\bf 4.} In this section we consider the in-medium correlation function of
the vector currents. First, consider the case of the isovector-vector
currents. The general 
expression for this correlator can be written as follows
\begin{equation}
\Pi_{\mu\nu}(p)=
i\!\int d^4x e^{ip\cdot x}\langle \Psi|T \{J_\mu(x)J_\nu(0)\}
|\Psi\rangle.
\label{correl}
\end{equation}
Where $J_\mu$ is the isovector vector interpolating current.
The lowest pole of this correlator corresponds to the $\rho$-meson
contribution. The problem of how medium with nonzero density and/or
 temperature  alters the $\rho$-meson properties compared to those in vacuum
attracts much attention nowadays \cite {Wa}.
Due to relatively large width it decays basically inside the nuclear interior
so the spectrum of the produced dileptons can carry, at least in principle,
 direct information about the modifications of the 
 $\rho$-meson mass and width in matter. Such modification may be related
to partial restoration of CS. The scaling relations proposed
in \cite{Br} suggested that the  $\rho$-meson mass should decrease in matter
following the behavior of the chiral order parameter. First calculations,
based on QCD sum rules \cite{Ha} indeed supported this conjecture and
 predicted a drop of the   $\rho$-meson mass by 20$\%$ at 
 normal nuclear density. One notes that in Ref.\cite{Ha} the nuclear medium
was represented as a system of  noninteracting nucleons and for the 
phenomenological spectral density the pole-plus-continuum anzats was used.
Besides, the 4-quark condensate was approximated by the corresponding
factorized expression, which, as one could see in the previous section,
violates the chiral symmetry constraints. In more recent
calculations \cite {We1} the phenomenological spectral density
was treated in a more realistic way taking into account
$\rho N$ scattering process. These calculations demonstrated that a more 
realistic model of the spectral density leads to much less pronounced
alternations of the  $\rho$-meson mass compared to the pole-plus-continuum model.
Moreover, some phenomenological models  predict even the increase
of  $\rho$-meson mass in matter \cite{Pi}.
 One can thus conclude that this issue
is not completely settled. The other, at least theoretical, way to look
at the phenomena of CS restoration is to study the 
 correlators describing the in-medium dynamics of the chiral partners.
As we have already mentioned the correlators of the chiral partners,
in our case the correlators of the vector and axial-vector currents, should
become identical in the chirally restored phase. Thus one can expect that
these correlators get mixed when the symmetry is only partially restored.
In other words the difference between the 
correlators of the chiral partners, being in some sense an order parameter,
tends to become smaller with increasing density and/or 
temperature. 
The  effect of chiral mixing indeed takes place both at finite
 temperatures \cite{De}
and densities \cite{Kr}. We note that in the case of nuclear density
with zero temperature the ``axial phase'' of the vector correlator
should be accompanied by particle-hole excitation of 
the nucleus to preserve the total isospin. This point reflects the 
difference between the cases of finite temperature and finite density
although it should rather be viewed as a ``pure nuclear'' effect.
From the CS point of view the chiral mixing is the manifestation of the same
physics both at finite $T$ and $\rho$.
 We note that in the real case 
of the heavy ion collisions the chiral mixing may happen without such
particle-hole intermediate excitation. 
The axial vector correlator contains the 
contribution from pion and axial-vector meson $A_1$ which is a chiral partner 
of the $\rho$-meson. Being a Goldstone
boson, pion  experiences little change in matter 
up to the point of the chiral phase transition. Pion decay constant $F_\pi$ scales
with density like the order parameter and can be fixed relatively well up to
a density close to normal nuclear one. It is much more difficult to
describe the in-medium properties of the  $A_1$ meson. This meson can decay 
into three pions, each of them can interact with nuclear medium and
with each other. All this makes it very difficult to estimate the 
correlator of the axial-vector current with sufficient accuracy. This
in turn significantly increases the size of uncertainties related to the 
calculation of the in-medium mass of the $\rho$-meson since, due to 
mixing property, the vector correlator 
in nuclear medium acquires an additional singularity
 describing  the contribution of the  $A_1$  meson which 
is multiplied by the corresponding, in general unknown, residue.
One notes that we discuss the idealized case of the $\rho$ meson at rest.
In reality they move and it may result in significant effect \cite {Lee1,Mo}
somehow relaxing the constraints following from CS. Making
 use of the standard commutation relation of current algebra
$\left[Q_5^a, J_\nu^b\right]=i\epsilon^{abc}A^{c}_\nu$
and putting it in the general expression for the correlator of the vector currents
$\Pi_V$
one can get 
\begin{equation}
\Pi_V=\overcirc\Pi_V+\xi(\overcirc\Pi_A-\overcirc\Pi_V)
\label{PiV}
\end{equation}
and similar expression for the axial vector correlator $\Pi_A$
\begin{equation}
\Pi_A=\overcirc\Pi_A+\xi(\overcirc\Pi_V-\overcirc\Pi_A)
\end{equation}
Here  we defined
\begin{equation}
 \xi={4\rho\overline{\sigma}_{\pi{\scriptscriptstyle N}}\over 3f_\pi^2 m_\pi^2}.
\end{equation}
 and denoted $\overcirc\Pi_V$($\overcirc\Pi_A$)
the correlator of the vector (axial) currents
calculated in the approximation of the noninteracting nucleons with finite
density.
As one can see from the above equations the correlators get mixed when soft
pion contributions  are taken into account. We note that this statement
is model independent and follows solely from CS. In contrast,
 studies of the $\rho$ meson in medium done so far focused on the 
calculations of different terms of the $\rho$ meson self energy sometimes
using some model dependent assumptions. On the other hand
 much less attention has
been paid on the general chiral structure of the corresponding correlators.  
It is worth emphasizing that CS alone cannot predict the actual
behavior of the in-medium $\rho$ meson mass. Instead, chiral symmetry implies
that the correlators of the chiral partners acquire, due to mixing,  
additional singularities. It is an additional pole when the soft pion
 approximation is used, but in the general case of pions with  nonzero 
3-momentum the axial correlator exhibits a cut.
These singularities may manifest themselves in the dilepton 
spectrum, produced in the heavy-ion collisions. It means that the spectrum may
 show an additional enhancement in the energy region close to the mass
of the $A_1$ meson, besides that at the mass of   $\rho$ meson. It is interesting 
to note that in this case pions
act against chiral symmetry restoration unlike the corresponding mixing
relation at finite temperature.\\
One notes that the $\omega$ meson being an isospin singlet is not affected
by the chiral corrections. It follows from the fact that the commutator
of the isoscalar vector current with the axial charge $Q_a^5$ is zero.
That in turn means that the    $\omega$ meson case is in some sense
cleaner than the one with   $\rho$ meson since there is no chiral mixing.
Of course,  $\omega$ meson in medium is affected by many other
nuclear effects but at least this source of uncertainty is absent.
Now  a few remarks concerning the general case
of matter with nonzero density and temperature are in order . This regime
 is the one 
which is realized in  genuine heavy ion collisions. Since from the
CS point of view a system with finite temperature does not
differ from  one with finite density it is natural to expect 
that the phenomena of chiral mixing exists in a medium with finite
temperature and density with separate contributions from the virtual
nuclear and thermal pions. One could also expect some sort of mixed
contribution when pions, being emitted by the thermal bath, are then absorbed by
the nuclear medium. The analytical 
structure of the corresponding correlators becomes more complicated
and may contain singularities due to both thermal and nuclear 
pions.
These and related questions will be the subject of a
forthcoming publication.\\
{\bf 5.} In this section we briefly address the issue of the chiral corrections
to the lowest order result for the scalar in-medium two-quark condensate
given by Eq.(15).
Let's first illustrate the derivation of the lowest order result.
Assuming that the condensate at finite density can be represented as the 
sum of the vacuum contribution and the contribution coming from the system
of the noninteracting nucleons with constant density one can write 
\begin{eqnarray}
\langle A|\overline {q}q|A\rangle_\rho =
\int d^3x \langle A|\overline {q}q(x)|A\rangle 
=\langle 0|\overline {q}q|0\rangle +\rho\int 
d^3x{\langle N|\overline {q}q(x)|N \rangle}\nonumber\\
=\langle 0|\overline {q}q|0\rangle 
 (1-{\sigma_{\pi{\scriptscriptstyle N}}\over f_\pi^2 m_\pi^2}\rho),
\end{eqnarray}
Here we have used the definition of the $\pi N$ sigma term and GOM relation. 
The corrections to this lowest order expression were studied in a number 
of papers \cite {We,Ko}. They arise when
 nuclear interactions mediated by mesons are taken into account.
  One notes that, while the lowest order
 expression is model independent, the higher order contributions inevitably
call for model dependent assumptions. 
In this section we are specifically interested in corrections which 
solely come from CS and,
in some sense, are model independent . They can be obtained if 
the T-product of two interpolating fields in Eq. (\ref{correl}) is replaced
by the operator $\overline {q}q$. Following the procedure outlined above
and calculating corresponding commutators one can get 
\begin{equation}
\langle A|\overline {q}q|A\rangle _\rho =\langle 0|\overline {q}q|0\rangle (1-
{\rho\sigma_{\pi{\scriptscriptstyle N}} \over f_\pi^2 m_\pi^2}
(1-
{\rho\overline{\sigma}_{\pi{\scriptscriptstyle N}} \over f_\pi^2 m_\pi^2})),
\end{equation}
One can see from the  above expression that pion corrections result in
 an increase of the overall correction to the vacuum value of the two quark
condensate by 12-15$\%$ depending on the value accepted for 
$\overline{\sigma_{\pi{\scriptscriptstyle N}}}$. In the other terms, the leading
chiral corrections to the two quark condensate are entirely governed by the 
long-ranged part on the nuclear pion field.  
As was earlier mentioned
the soft pion contribution to the in-medium two-quark condensate is not 
related to the chiral symmetry restoration but it is always useful
to estimate the contribution of such terms since it gives a qualitative
idea about the relative size of the 
corrections which are not due to soft pions and could thus
can be relevant when the symmetry restoration related problem
is considered. Now a remark concerning the issue of the two
 quark condensate
in finite nuclei is in order. In finite nuclei the density is no longer
a constant but a function of the nuclear radius. Thus the two quark condensate
effectively becomes nonlocal. Besides, this ``nuclear'' nonlocality
, there is also ``vacuum'' nonlocality stemming from the fact that the 
standard two quark condensate is the first term of the short-range expansion
of the nonlocal expression  $\overline {q}q(x)$. Both nonlocalities should
, in principle be included. However, the scales of these two 
types of nonlocalities are different and they are not related to 
each other. Moreover, in the case of finite nuclei the ``nuclear'' 
nonlocality can be accounted for by the local
density approximation. One notes, that in the processes where
time plays an important role as in heavy ion collisions, the time dependence of the 
quark condensate should also be included \cite{Fr}. Thus we see  that in the real
situation the in-medium quark condensate becomes model dependent even in the 
lowest order in  nuclear density.\\  
{\bf 6.} In this section we briefly consider the mixing of the other type of chiral
partners, namely the $\sigma$-$\pi$ pair. The corresponding relations can
 be obtained straightforwardly from the general expression of  the current-current
correlator.  Making use of the current algebra commutation relation
\begin{equation}
[Q_5^a, \pi^b]=-\delta^{ab}i\sigma.
\end{equation}
and
\begin{equation}
[Q_5^a, \sigma]=-i\pi^a.
\end{equation} 
one can get the mixing relations
\begin{equation}
\Pi_\pi=(1+\xi)\overcirc\Pi_\pi-\xi(\overcirc\Pi_\sigma)
\label{PiPi}
\end{equation}
and 
\begin{equation}
\Pi_\sigma=(1+\xi)\overcirc\Pi_\sigma-\xi(\overcirc\Pi_\pi)
\label{PiA}
\end{equation}
Here  the mixing parameter is
\begin{equation}
 \xi={2\rho\overline{\sigma}_{\pi{\scriptscriptstyle N}}\over 3f_\pi^2 m_\pi^2}.
\end{equation}
One can see that the effect of $\sigma$-$\pi$ mixing is less pronounced at the normal 
nuclear density than in the case of $\rho$-$A_1$ mixing. The point of the complete
restoration of chiral symmetry should, of course, be the same for all kinds of
the chiral partners but the ``velocity'' of approaching to this point 
may well be 
different. We note that via the mixing with $\sigma$ the pion polarization operator
acquires an additional width. Similar to the case of the $\rho$-$A_1$ system the mixing in the 
pseudoscalar-scalar channel practically means that the correlators exhibit the singularities
which are dictated by CS and should be taken into account regardless of the model used to
describe the concrete hadronic processes.  The phenomena of mixing suggests
a few possible ways of how chiral symmetry restoration occurs. One notes
that these scenarios are similar to those found earlier for the case of 
finite temperatures \cite{Shu}. Firstly,
 the in-medium masses or, equivalently, 
the lowest singularities of the corresponding correlators may become the
 same at the point of restoration. Secondly, both correlators may exhibit
two poles of the same strength corresponding to different mesons. 
In this scenario the identity of the correlators does not lead to the  
same masses of the chiral partners at the point of restoration. In these
two scenarios it is implicitly assumed that the mesons still retain their 
individuality even at high densities. In other words, the in-medium widths
 are small compared to the mass difference
of the mesons. One could suggest the other scenario where the widths are
of the order of  mass  difference so that the spectral
 functions get smeared over the wide energy region. Thus
it no longer makes sense to discuss the individual quantum states with
the mesonic quantum number when the density is high enough. One needs to mention that the issue
of the experimental study of the CS restoration using the chiral
mixing relations is a completely open question so neither of 
the three scenarios can be ruled out. 
 The effect of the $\sigma$-$\pi$ mixing can probably be observed at 
relatively large densities. The case worth studying is deeply bound pion states
\cite {Kk,Ya} in heavy nuclei recently observed in $(d,^{3}He)$ reaction \cite {Ya1}.
The other possible method to observe the effect of mixing and closely related phenomena of
CS restoration is to measure the process of two pion production in hadron-nucleus
interactions. Such kind of experiment has recently been made \cite{Eba} on a number of
nuclei ranging from deuteron and up to heavy targets. The results of the measurements
indicate the strong enhancement of the $\pi^+\pi^-$ pairs yield (in I=J=0 state) 
with  increase of the nuclear mass number. One notes that the results of a recent
theoretical paper \cite{Ha1} support the idea that the large  enhancement of the  $\pi^+\pi^-$
production near threshold is due to partial restoration of CS. The $\sigma$-$\pi$ mixing
, considered in the present paper may well be  responsible for the significant part of 
this enhancement.\\
{\bf 7.} In this section we discuss the chiral corrections to the 3-point
correlation function. The studies of hadrons in nuclear matter are mainly
focused on the in-medium 2-point correlators responsible for such
effects as mass shifts. There are only a few calculations of the in-medium properties 
of the 3-point functions (see for example Refs.\cite{Br,Koch}). However the 
careful treatment of the corresponding medium corrections is an
inevitable element of any consistent theory of nuclear interactions since
the modifications  experienced by the correlators may lead to 
observable effects in particle spectrum \cite{Koch}. Even considering the $\rho$ meson mass
shift it may become important to provide the $\rho\pi\pi$ vertex function satisfying
the chiral symmetry constraints. The in-medium 3-point correlation function has the following
general form 
\begin{equation}
\Pi(p,q)=
i\!\int d^4x d^4y e^{ip\cdot x}e^{iq\cdot y}\langle \Psi|T \{J_1(x)J_2(y)J_3(0\}
|\Psi\rangle.
\end{equation}
where $J_{1(2,3)}$ is the hadron interpolating current. One notes that carring out the standard
QCD SR strategy is extremely complicated in this case. One of the relative distances may become
small making the  standard OPE impossible to use.\\
Let's denote by $V, P, S, A$ the vector-isovector, pseudoscalar, scalar and axialvector
interpolating currents and consider the VPP 3-point correlation function. This correlator appears
when the in-medium modifications of the  $\rho\pi\pi$ coupling constant are studied.
Making use of the current algebra commutation relations one can get the 
following mixing property for the PPV 3-point correlation function

\begin{equation}
\Pi_{PPV}(p,q)= \Pi_{PPV}^{0}(p,q) + \xi(\alpha\Pi_{PPV}^{0}(p,q) + \beta\Pi_{SSV}^{0}(p,q)
+ \gamma\Pi_{SPA}^{0}(p,q))
\end{equation}

Here the coefficients $\alpha, \beta$ and $\gamma$ depend on the isospin indices. 
 The mixing parameter is
\begin{equation}
 \xi = {\rho\overline{\sigma}_{\pi{\scriptscriptstyle N}}\over f_\pi^2 m_\pi^2}.
\end{equation}
As in the case of 2-point functions we denote by $\Pi_{SSV}^{0}$ the correlator calculated without mesonic
 corrections. As one can see all possible chiral partners enter and it makes the computation 
of the corresponding observables rather involved. A similar relation holds for this 3-point function
considered in the medium with finite temperature.  One notes that the analogous mixing
property can be obtained when, instead of the vector-isovector current, the vector-isoscalar
one is considered. In this case the mixing looks simpler due to the zero commutator of
the axial generator with the vector-isoscalar current.  

\begin {center}
\bf Conclusion
\end{center}
We considered the relationships between chiral symmetry and
properties of the in-medium hadron correlators. It was demonstrated that chiral
symmetry imposes  important constraints on the correlators. In the case 
of the nucleon correlator it does not allow  terms linear in pion mass
to appear in the expression for the nucleon mass. It also fixes  certain
properties of the pieces of the nucleon correlator with  different Dirac
structures. The other important consequence of the chiral symmetry is that 
the factorization approximation, sometimes used to parameterize the in medium 
four quark condensate is not valid. In the case of the mesonic  
chiral partners such as  $\sigma$-$\pi$ and $\rho$-$A_1$ pairs chiral symmetry
results in the mixing of the corresponding correlators similar to those
established earlier for the case of finite temperatures. 
This gives rise to  additional singularities of the correlators
and may lead to the enhancement of experimental spectra in threshold region.
This mixing can
be viewed as an indication toward the restoration of chiral symmetry.
 It also suggests several possible scenarios of such a 
restoration. We also considered the soft-pion corrections to the
in-medium scalar two-quark condensate and discussed the chiral mixing property for the case
of the 3-point functions.
  
\section*{Acknowledgments}

The author wants to thank Mike Birse for  enlightening discussions concerning
the role of chiral symmetry in nuclei. Useful discussions with T.Cohen,
L.Kisslinger, T.Hatsuda, M. Johnson and A.Thomas are gratefully acknowledged.
The author is grateful for the support from Center for 
Subatomic Structure of Matter at the University of Adelaide where the final part of this 
work was done.

\end{document}